\documentclass[12pt]{article}

\usepackage{graphicx}
\usepackage{amssymb}
\usepackage{epstopdf}
\usepackage{amsfonts}
\usepackage{amsmath}
\usepackage{bm}
\usepackage{mathrsfs}
\usepackage{subfigure}

\def\be{\begin{equation}}
\def\te{\end{equation}}
\def\ee{\end{equation}}
\def\nn{\nonumber}
\def\bea{\begin{eqnarray}}
\def\tea{\end{eqnarray}}
\def\eea{\end{eqnarray}}

\begin{document} 
\title{\Large Quantum Hierarchical Systems: Fluctuation Force by Coarse-Graining, Decoherence by Correlation Noise} 
\author{\large Bei-Lok Hu}
\date{\small University of Maryland, College Park, USA (blhu@umd.edu) August 21, 2020}
\maketitle

-- {\it Dedicated to the memory of Professor Heinz Dieter Zeh who made invaluable contributions to the foundational issues of quantum mechanics \cite{ZehFQM} and pioneered the environment-induced quantum decoherence program \cite{ZehDec}.}

\begin{abstract}
While the issues of dissipation, fluctuations, noise and decoherence in open quantum systems (with autocratic divide) analyzed via Langevin dynamics are familiar subjects, the treatment of corresponding issues in closed quantum systems is more subtle, as witnessed by Boltzmann's explanation of dissipation in a macroscopic system made up of many equal constituents (a democratic system). How to extract useful physical information about a closed democratic system with no obvious ways to distinguish one constituent from another, nor the existence of  conservation laws governing certain special kinds of variables, e.g., the hydrodynamic variables -- this is the question we raise in this essay.    
Taking the inspirations from Boltzmann and Langevin, we study a) how  a \textit{hierarchical order} introduced to a closed democratic system -- defined either \textit{by substance or by representation}, and b) how   \textit{hierarchical coarse-graining}, executed in a specific order, can facilitate our understanding in how macro-behaviors arise from micro-dynamics. We give two examples in: a) the derivation of correlation noises in the BBGKY hierarchy and how using a Boltzmann-Langevin equation one can study the decoherence of the lower order correlations; and b) the derivation of quantum fluctuation forces by ordered coarse-grainings of the relevant variables in the medium, the quantum field and the internal degrees of freedom of an atom. 
\end{abstract}

-- To appear in the H. D. Zeh memorial volume of the Springer Series on ``Fundamental Theories of Physics", edited by C. Kiefer (2021).


\section{\large Democratic, Autocratic and Hierarchical Systems}

A more familiar set of nomenclature is, respectively, ``Closed, Open and Effectively Open" systems. These two sets differ in their characterization but there are correspondences between them.  The title set refers to  the system's constituents: democratic means all constituents are equal, autocratic means there is a distinguished  party (or a few in the case of oligarchic) and the rest are irrelevant (the `peasants').  The latter nomenclature refers to the degree the system is influenced by its environment: a closed system has no environment, an open system has \cite{qos}. In an autocratic system, it is easy to identify the chosen (few) as the distinguished and the rest acting as its environment, in the sense that the distinguished does not care about the details of the `irrelevant' individuals, perhaps only their collective opinion in a very coarse way, as noise
(maybe not completely white, but somewhat colored).  Thus in this sense an autocratic system can easily be dealt with in the manner of an open system and the distinguished party's behavior described by the Langevin dynamics.  For a closed  democratic system, all constituents are identical. Consider for example a mole of monatomic gas.   The question is, how to describe the macroscopic features of this gas, made up of a huge number of identical particles, from the microscopic dynamics of each of these particles, knowing that the micro and macro behaviors are fundamentally different. Even possessing the details about the positions and momenta of each particle one cannot straightforwardly predict the dissipative macro-dynamics and the appearance of an `arrow of time' from the unitary micro-dynamics of these particles. 

This is a daunting challenge  Boltzmann took upon himself and resolved with brilliant ideas: First, introduce a hierarchical representation of the nth order correlation functions of the molecules.
Then focus on a lower order correlation function -- the lowest being the one-particle distribution function. It is driven by the higher order correlation functions -- the lowest being the two point function, which gives a reasonable description of a dilute gas.  Now introduce the molecular chaos assumption, namely, assume that the colliding partners are uncorrelated initially -- we shall refer to this `causal factorization condition' as `slaving'.   
This leads to the Boltzmann equation which depicts the dissipative dynamics of a system of (dilute) molecular gas. 

Using  a hierarchical representation of this closed system and  focusing attention on a certain sector (the lower order correlations) while including the influence of another sector (the higher order correlations) under certain physical conditions, we can treat this otherwise closed system in an open system manner.  It is in this sense that we can call it an `effectively open' system.  How to extract useful information about a closed democratic system with no obvious way to distinguish one constituent from another, nor are there conservation laws governing certain kind of special variables (e.g., the hydrodynamic variables) -- this is the question we pose in this essay, a study of closed quantum systems by hierarchical representations.  With the inspiration of Boltzmann and Langevin, we shall study how  a hierarchical order introduced to a closed democratic system -- defined either by substance or by representation --  can indeed facilitate our understanding of how macro-behaviors arise from micro-dynamics. 

The remainder of this section summarizes  the two major paradigms of NEq statistical mechanics, a la Langevin and Boltzmann (see Chapter 1 of \cite{CH08}), and the key ideas behind open quantum systems: how persistent  structures emerge  from judicious choices of coarse-graining and non-Markovianity from self-consistent backreaction.   This will set the stage for the central themes of this paper, a hierarchical representation of a closed system rendering it an effectively open system,  and hierarchical coarse-graining in understanding natural phenomena in a systematic way. 
  
In the hierarchical conceptual framework, we shall give two examples showing how ordered coarse-graining in a systematic manner offers ways to 1) calculate the quantum  fluctuation force \cite{QFlucF}, the Casimir-Polder force \cite{CasPol} in our example,  between an atom and a dielectric material; 2) derive the effective noise from quantum fluctuations of the higher order correlations in the Schwinger-Dyson hierarchy of equations for an interacting quantum field theory which can act to decohere the  lower order correlations. Sec. II and III are about closed quantum systems, and Sec. V is devoted to this second task.  Beyond the questions of dissipation and time-arrow Boltzmann asked of closed classical systems, an equally fundamental issue is, how does classical behavior appear in a closed quantum system? This issue, often referred to as  decoherence \cite{DecohBooks}, has been explored extensively from the 80s and 90s under the consistent/decoherent history conceptual framework and the environment-induced decoherence paradigm. To motivate these issues we shall focus in Sec. II on the special roles of the collective and hydrodynamics variables. Sec. III discusses hierarchical systems in both structural and representational senses. We show their particularly adapt characteristics for addressing these issues in closed quantum systems.  The derivation of the correlation noise from the nPI (master) effective action is left to Sec. V.  

Section IV introduces hierarchical coarse-graining (CG) and discusses how, for the same set of variables or components in a closed system,  different ordering of CG yields different results. We use the calculation of a fluctuation force, the Casimir-Polder force, as example. In the atom-field-medium system of interest we see 5 variables associated with the three parties involved: the atom has two and the medium has two. The primary action is between the atom's internal degree of freedom (dof) interacting with a quantum field which is modified by the presence of a dielectric. We show how an orderly coarse-graining over these 
5 variables (see Table I) by means of the \textit{graded influence action} yields the Casimir-Polder force.  We use this exactly controllable example with unambiguous results to illustrate the idea of hierarchical coarse-graining\footnote{Looking up the web, I found this terminology  appears most often in relation to biological systems. Ideas akin to this, likely broader in scope and in variety, for the description of real and perceived structures in the physical world and adaptive systems can be found in this interesting paper \cite{GMH14} } which has deeper implications and broader scope of applications.

\subsection{Open systems: coarse-graining and
backreaction} 

The first order of business in treating an open system is to \textit{identify} which subsystem or which variables (the relevant sector) are of special interest and whether the remainder (the irrelevant sector) constitutes an environment whose details are  of lesser  importance, and only a coarse-grained description is sufficient to summarize its overall effects on the system of special interest.  To make a sensible distinction involves recognizing and devising a set of criteria to \textit{separate} the relevant from the irrelevant variables. This procedure is simplified when the two sets of variables possess very different characteristic time or length or
energy scales or interaction strengths. An example is the separation of slow-fast or heavy-light variables as in the Born-Oppenheimer approximation in molecular physics where the nuclear variables are assumed to enter adiabatically as parameters in the electronic wave function. Similar separation is possible in quantum cosmology between the `heavy' gravitational sector characterized by the Planck mass and the
`light' matter sector \cite{Kiefer}. In statistical physics this separation can be made formally with projection operator techniques \cite{ProjOp}. This usually results in a nonlinear integro-differential equation for the relevant variables, which contains the causal and correlational information from their interaction with the irrelevant variables.

Apart from finding some way of {separating} the overall closed system into a `relevant' part of primary
interest (the open system) and an `irrelevant' part of secondary interest (the environment) in order to render calculations possible, one also needs to devise some {\it averaging} scheme to reduce or reconstitute the detailed information  of the environment such that
its effect on the system can be represented by some macroscopic functions, such as the transport or response functions. This involves introducing certain {\bf coarse-graining} measures.  It is usually by the imposition of such measures to varying degrees that an environment is turned into a bath, and certain macroscopic characteristics such as temperature and chemical potential can be introduced to simplify its description.  

How the environment affects the open system is determined by the {\bf back-reaction} effects. By referring to an effect as backreaction, it is implicitly assumed that a system of interest is preferentially identified, that one cares much less about the details of the other sector (the `irrelevant' variables in the `environment'). The backreaction can be significant, but should not be too overpowering,
so as to invalidate the separation scheme.  To what extent one views
the interplay of the two sectors as {\it  interaction} (between two
subsystems of approximately equal weight) or as {\it backreaction}
(of a less relevant environment on the more important system) is
reflexive of and determined by the degree one decides to keep or
discard the information in one subsystem versus the other.  It also
depends on their interaction strength. Through reaction and
backreaction the behavior of each sector is linked to the other in an inseparable way, i.e., by their interplay.

{\bf  Self-consistency} is thus a necessary requirement in
backreaction considerations. This condition is imbued in the fluctuation-dissipation theorem (FDT).
When the environment is a bath,  for systems near
equilibrium, their response can be depicted by  linear response
theory. Even though such relations are usually presented in such a
context,  its existence in a more general form can be shown to cover
nonequilibrium systems. Indeed as long as backreaction is included,
such a relation can be understood as a corollary of the
self-consistency requirement, which ultimately can be traced to the
unitary condition of the original closed system.

A familiar example of self-consistent backreaction process is the
time-dependent Hartree-Fock approximation in atomic physics or
nuclear physics, where the system could be described by the wave
function of the electrons obeying the 
Schr\"odinger equations with a
potential determined by the charge density of the electrons
themselves via the Poisson equation. In a cosmological backreaction
problem, one can view  the system as a classical
spacetime, whose dynamics is determined by Einstein's equations with
sources given by particles produced by the vacuum excited by the
dynamics of curved spacetime and depicted by the appropriate wave equations in this spacetime.

Much of the physics of open systems is concerned with the
appropriateness in the devising and the implementation of these procedures, namely, {\it separation, coarse-graining} and {\it backreaction}.  1) the identification and separation of the
physically interesting variables which make up the open system, 2) the `averaging' over of the environment or the projecting out of the irrelevant variables -- how different coarse-graining measures
affect the final result is important (we give an example in Sec. IV), and 3) the evaluation of the averaged effect -- the backreaction -- of the 
environment on the system of interest. 

These considerations surrounding an open system are common and
essential not only to well-posed and well-studied examples of
many-body systems like molecular, nuclear and condensed-matter
physics, they also bear on some basic issues at the foundations
of quantum mechanics and statistical mechanics, such as
decoherence and the emergence of classical behaviors,
with profound implications on the emergence of time, spacetime, even quantum mechanics as an emergent  theory \cite{Adl04,tHemQM}.

\subsection{From closed to effectively open systems} 

There are many systems in nature which are apparently closed (to the observer), in that there is no obvious way to define a system which is so much different from an environment. These systems  do not possess any parameter which can enable the observer to distinguish possible heavy-light sectors, high-low frequency behavior or slow-fast dynamics. Boltzmann's theory of molecular
gas is a simple good example: All molecules in the gas are on equal footing, in that no one can claim to be more special than the others.
Because of the lack of separation parameters which marks the discrepancy of one component from the other, these systems do not lend themselves
to an obvious or explicit system - environment divide like open systems would, and appear like closed.  However,  usually in their effective description a separation is introduced
implicitly or operationally because of their restricted appearance or due to the imprecision in one's measurement. These are called effectively open systems. Boltzmann's treatment of molecular gas, how he ``effectively opened up" a closed democratic system,  is a sterling example we shall follow to address a number of important issues in closed quantum systems. 

\subsection{Paradigms of nonequilibrium statistical mechanics}

We can highlight the distinction between open and effectively open systems by comparing the two primary models which characterize these
two major paradigms of  nonequilibrium statistical mechanics: the Boltzmann-
{Bogoliubov-Born-Green-Kirkwood-Yvon (BBGKY) hierarchy} theory \cite{Balescu}
of molecular kinetics, and the Langevin
(Einstein-Smoluchowski) theory \cite{vanKamp}
of Brownian motions. The difference between the two are of both formal and conceptual nature.

To begin with, the {\em setup} of the problem is different: As we remarked above, in kinetic theory one studies the overall dynamics of a system of gas molecules, treating each molecule in the system
on the same footing, while in Brownian motion one (Brownian) particle which defines the system is distinct, the rest is relegated to be the environment. The terminology of `relevant' versus
`irrelevant' variables not so subtly reflects the discrepancy.

The {\em object} of interest in kinetic theory is the one-particle distribution function (or the nth-order correlation function), while in Brownian motion it is the reduced density matrix. The emphasis in the former is the behavior of the
gas as a whole (e.g., dissipative dynamics) taking into account the correlations amongst the particles, while in the latter is the motion of the Brownian particle under the influence of its
environment.

The nature of {\it coarse-graining} is also very different: In Brownian motion it is in the
integration over the environmental variables. In kinetic theory coarse-graining is a bit more subtle. It resides in confining ones attention to one particle distribution functions, assuming a factorization condition for the two point functions under the molecular chaos assumption. This corresponds formally to a) a truncation of the BBGKY hierarchy,  and b) an introduction of a causal factorization condition, the combination of these two steps  we refer to as ''slaving'', the pivotal process which renders a closed system effectively open.  Note that truncation would only yield two  disjoint partitions, each one being a smaller closed system separated from the complete hierarchy. The part that is largely `ignored' is where the noise comes from, while its main physical effect on the `system' is to impart a dissipative component in its dynamics and for a quantum system, acting as the source of decoherence \cite{GelHar93}. 

Finally the {\it philosophies} behind these two paradigms are quite different: In problems which can be modeled by Brownian motion, the separation of the system from the environment is prescribed: it is usually determined by some clear disparity between the two systems. In closed systems like a molecular gas where there is no clear discrepancy in sight, making a convenient yet physically unjustified separation often leads to wrong predictions. 
Coarse-graining in Boltzmann's kinetic theory is also very different from that of Brownian
motion. The latter is explicit and thus easily identified while the former is more subtle, as the system has its own innate systematics \cite{ZehQcor}. However, as we shall see in Sec V, one effective way of coarse-graining a la Boltzmann lies in the `slaving' procedures, where information attached to higher correlation orders is not kept in full, but represented by the so-called `correlation noise', which turns a Boltzmann equation into a stochastic Boltzmann equation \cite{StoBol}. Now, just what correlation order is sufficient for the physics under study is an objectively definable and verifiable act, which ultimately is determined by the degree of precision in a measurement and judged
by how good it depicts the relevant physics.


\section{\large Collective and Hydrodynamic Variables}

In treating physical systems containing many micro degrees of freedom a meaningful challenge is to design a small set of meso or macro variables to render the problem at the micro level technically tractable while preserving its physical essence. Instead of variables one can also seek to reformulate the theory such that in the domain of interest, e.g., at low energy, an effective theory gives an accurate enough description.  In order to construct  a simple and accurate enough theory valid at a specified range (of mass, energy, length or time scales) the challenge comes first and foremost in the design of such sets of  variables.    Familiar examples abound, e.g., thermodynamics from statistical mechanics, hydrodynamic limit of kinetic theory, collective dynamics in condensed matter and nuclear physics.
Bear in mind that a coarse-grained description in terms of thermodynamic or hydrodynamic variables of a collective nature and their associated response functions  is qualitatively very different from the detailed description  in terms of the underlying microscopic variables and dynamics. E.g., the notion of temperature and pressure in thermodynamics makes no sense for each molecule.   A theory of collective variables  which emerges from the microscopic theories  may  bear little resemblance to them.   

We comment on two such kinds of variables and the effective theories they bring out, namely, the collective and the hydrodynamic variables.
We begin with the collective variables which are general and  broadly used, we then look at a special class, the hydrodynamic variables, which has been studied in great detail in the 90s in the context of consistent / decoherent histories.   Because of the special properties they possess, namely, obeying conservation laws, they have a clear advantage in becoming `habitually' decohered in the quasi-classical domains which emerge.

\subsection{Collective Variables}

Collective variables are derived variables, in the sense that they are not elemental (even quarks and gluons may one day be  shown to be composites), and the theories these variables describe are effective theories.  When we refer to one branch of physics as particle  physics and some other branch of physics as molecular physics, we are already invoking the concepts of collective variables. In studying molecular forces we rarely need to include quark-gluon processes, much less string theory. Molecular theories are effective theories constructed from atomic wavefunctions. Knowing QED and the atomic composition of a molecule helps to construct viable molecular theories for the properties of molecular forces and the dynamics of molecular interactions.  Deriving nuclear forces from QCD faces the same kind but more difficult challenge. 

In order to construct a certain set of collective variables to serve a specific purpose, one often needs to have some idea in what type of physics are of special interest, and in what parameter ranges. E.g., the many `ons' -- phonon, roton, plasmon etc -- in condensed matter physics are collective excitations, thus it is the quantized normal modes of vibration or rotation of many atoms or the excitation modes of a plasma. Distinguishing these specific excitation modes by different collective variables serves targeted functions. These variables obey equations of motion or possess symmetry properties very different from that of the basic constituents (atoms) they are constructed from. In fact, the collective physics they describe often ceases to make sense when one examines it at the atomic scale. This is true of the hydrodynamic variables we shall discuss in the next subsection.         

Consider one very common type of collective variable, the  {center of mass} (CoM) of an assembly of particles. It is defined in such a way that when acted upon by some external force the CoM of a system of particles with total mass M moves like a single particle of mass M. It is a collective variable,  not for the description of some type of collective excitation, but as a configuration variable constructed to simplify the  description of the dynamics of the collective. 
Let us examine the role this CoM variable plays and the specific conditions which make  it special playing this role in two well-known processes: 1) decoherence of a dust particle in a gas -- why does it decohere so fast?  and 2) the special role of the CoM in representing the quantum properties of a macroscopic object. 


\subsubsection{Decoherence in a gas environment}

A quantum object collides with many particles in its ambient environment. We are interested in how fast this object decoheres and begins to behave classically. After the pioneering work of Joos and Zeh (in \cite{ZehDec}), many authors have studied this problem, e.g., \cite{GalFle,Dio,Hor,DodHal,Pol}.  
Take the calculation of Hornberger as example.
In calculating the S-matrix he made a reasonable assumption that, in general, a {collision keeps the center-of-mass invariant, and only the relative coordinates are affected}. Results become simplified in the limit $m/M \rightarrow 0$, namely, when the dust particle's mass M is much bigger than that m of the ambient particles.  
Putting in the typical mass of a dust particle in the  interstellar medium and as  background the cosmic microwave radiation, even assuming a very long relaxation time (the age of the universe) the width of a solitonic wave packet describing the center of mass is as small as 2 pm. He uses this example to show ``the remarkable efficiency of the decoherence mechanism to induce classical behavior in the quantum state of macroscopic objects". 
 
With the assumption that the mass M of the quantum object is much greater than m, that of the ambient particles, one can describe this object's dynamics as a kind of quantum Brownian motion, and view its decoherence in an open quantum system framework, as environment induced. It is of interest to note that this problem can be dealt with in the closed quantum system framework. No environment need be brought in to see the dust particle's decoherence. E.g., Dodd and Halliwell \cite{DodHal}  cast the master equation for the dust particle in a form  {emphasizing the role of local densities}. Being hydrodynamic variables they decohere habitually  because of the conservation laws they obey. More of this in the next subsection. Let us continue with how the collective variables can serve  the function of depicting macroscopic properties from the micro-constituents, even quantum properties. 

\subsubsection{Macroscopic Quantum Phenomena:  Center of Mass Axiom} 

What are the conditions upon which the quantum mechanical and statistical mechanical properties of a macroscopic object can be described adequately in terms of the kinematics and dynamics of its center-of-mass (CoM) variable?
This is an implicit assumption made in many macroscopic quantum phenomena (MQP) investigations, namely,
that the quantum mechanical behavior, such as quantum decoherence, fluctuations, dissipation and entanglement, of a macroscopic object, like the nanoelectromechanical oscillator, a mirror \cite{Marshall}, or a $C^{60}$ molecule \cite{Arndt}, placed in interaction with an environment can be captured adequately by its CoM behavior. For convenience we refer to this as the `CoM axiom'. This assertion is intuitively reasonable, as one might expect it to be true from normal- mode decompositions familiar in classical mechanics, but when the constituent particles interact with each other in addition to interacting with their common environment, it is not such a clear-cut result. We need to better define the conditions which make this possible and in most cases, not.

With the aim of assessing the validity of the CoM axiom the authors of \cite{CHY}  consider  a system modeled by $N$ harmonic oscillators  interacting with an environment consisting of $n$ harmonic oscillators  and derive  an
exact non-Markovian master equation for such a system in a environment with arbitrary spectral density and temperature. These authors outlined a procedure to find a canonical transformation  from the individual coordinates $(x_i, p_i )$ to the collective coordinates $(\tilde{X}_i, \tilde{P}_i ), i=1,...,N$ where $\tilde{X}_1,\tilde{P}_1$ are the center-of-mass coordinate and momentum respectively. They considered a  class of couplings between the system and the environment in the form $f(x_i)q_j$ (instead of the ordinarily
assumed $x_i q_j$) and examined if and when the CoM coordinate dynamics separates from the reduced variable dynamics. They note that if the function $f(x)$ has the property $\sum_{i=1}^{N} f(x_i) = \tilde{f}(\tilde{X}_1) +
g(\tilde{X}_2,...,\tilde{X}_N)$, for example $f(x) = x$ or $f(x) = x^2$, one can split the coupling between the system and environment into couplings containing the CoM coordinate and the relative coordinates. They proceed by tracing over the environmental degrees of freedom $q_i$ to obtain  the influence action which characterizes the effect of the environment on the system.

However, as the authors of \cite{CHY} emphasized, the coarse-graining made by tracing over the environmental variables $q_i$ does not necessarily lead to the separation of the CoM and the relative variables in the effective action. When they are mixed up and can no longer be written as the sum of these  two contributions, the form of the master equation will be no longer be so clean-cut -- the dynamics of both the relative variable and the center-of-mass variable will be mixed-in.

For this CoM axiom to hold what these authors  find is, at least for the quantum Brownian motion of a system of N harmonic oscillators in an n-oscillator bath, that the coupling between the system and the environment need be bi-linear, in the form $x_i q_j$, that the potential $V_{ij}(x_i - x_j)$ is independent of the center-of-mass coordinate.  In that case, one can say that the quantum evolution of a macroscopic object in a general environment is completely described by the dynamics of the center-of-mass canonical variables ($\tilde{X}_1,\tilde{P}_1$) obeying a non-Markovian master equation of the Hu-Paz-Zhang \cite{HPZ} type.

What is the role of collective variables in MQP?  At least for the N harmonic oscillator model it says: 1) For certain types of coupling the center of mass (CoM) variable of an object composed of a large number of constituents does play a role in capturing the collective behavior of this object
2) For other types of coupling the environment-induced quantum statistical properties of the system such as decoherence and entanglement are   generally more complicated. 

There is much to be explored in MQP. An interesting related issue is the {decoherence by internal degrees of freedom} \cite{BruMlo,HilMld,Flo98},
especially if the internal dynamics is chaotic \cite{ParKim}.


\subsection{Hydrodynamic variables}

In the chronicle of the development of theories of decoherence, it is well-known that Joos \& Zeh \cite{ZehDec} and Zurek \cite{Zur80} in the 80s  pioneered the environment-induced decoherence program which form the cornerstone for many important theories and experiments in quantum information and foundation developed in the following decade. What was lesser known is the theories of consistent histories of Griffith \cite{Gri} and Omnes \cite{Omn}, until Gell-Mann and Hartle \cite{GelHarDecHis} in the 90s greatly developed the decoherent histories formalism. The histories formulation of quantum mechanics is a fundamental advance: quantum histories are described by amplitudes, and they interfere. Only when they are decohered sufficiently can probabilities be used to describe a quasi-classical reality. The decoherent histories formulation of quantum mechanics is particularly adapt in treating closed quantum systems \cite{GelHarCloseQ}.

Following this vein and asking what conditions would be conducive to histories getting decohered,  comes another important discovery in the 90s:  hydrodynamical variables. Reason: they obey conservation laws. We shall give a quick summary of this because it is   important for understanding the appearance of quasi-classical domains in quantum closed systems. I'll mention in passing that the same principle, symmetry,  underlying the conservation laws, which singles out the hydrodynamic variables, is what allows for the existence of decoherence-free subspaces \cite{dfss} in quantum information, albeit the latter is in the space of quantum states.

\subsubsection{Hydrodynamic variables and conservation laws}

A group of papers addressing this issue appeared in the later half of the 90s \cite{HLM95,Hal98,Hal99}. 
In the context of the decoherent histories approach to quantum theory, Halliwell showed that a class of macroscopic configurations consisting of histories of local densities (number, momentum, energy) exhibits negligible interference. This follows from the close connection of the local densities with the corresponding exactly conserved (and thus exactly decoherent -- exact decoherence implies exact conservation) quantities (which commute with the Hamiltonian), and also from the observation that the eigenstates of local densities (averaged over a sufficiently large volume) remain approximate eigenstates under time evolution. Histories of these variables will be approximately decoherent and that their probabilities will be strongly peaked about hydrodynamic equations. 

Decoherence of hydrodynamic modes is considered by Calzetta and Hu \cite{CHhydro} using the influence action. They show that long wavelength hydrodynamic modes are more readily decohered. The result is independent of the details of the microscopic dynamics, and follows from general principles alone.

\subsubsection{Harmonic Chain}

The linear quantum oscillator chain is a solvable example where these principles and criteria can be tested out. 
Halliwell \cite{Hal03} showed that decoherence arises directly from the proximity of these variables to exactly conserved quantities  which are exactly decoherent, and \textit{not} from environmentally-induced decoherence.  He further explored the approach to local equilibrium and the subsequent emergence of hydrodynamic equations for the local densities. 

Another noteworthy aspect related to our central themes is coarse graining --  the difference in outcomes from coarse-graining a local group versus a nonlocal group. For a quantum chain   Brun and Hartle \cite{BruHar}
found that noise, decoherence, and computational
complexity favor locality over nonlocality for deterministic predictability.


\section{\large Hierarchical Systems}

Recall the challenge we posed: How to extract interesting physics  at the meso or macro scale from a closed quantum system of democratic microscopic constituents. As we mentioned the hydrodynamic variables stand out, aided by the conservation laws. A more familiar approach is mean field theory, which is often used as a first order approximation to the full theory. It assumes that each independent particle interacts with an effective potential depicting the averaged effects of all the other particles. Vlasov equation for a plasma is of this nature. Mean field theory respects democracy in a nice way, all particles are treated equally, each contributes equally to and experiences the same potential. The mean field dynamics remain time-reversal invariant. A mean field description of quantum field theory incorporates the effects of quantum fluctuations  through  renormalized interaction potential and renormalized coupling constants in the system.  However at this level of approximation the dynamics of the system is without dissipation and without noise. Thus it cannot help with the decoherence issue -- in fact the background field decomposition commonly used in field theory skips around this issue as it is often assumed to be classical without asking how that comes about (see, e.g., \cite{CH95prd}).

Because a closed democratic system has no environment, and no obvious scale differentiation, it is not easy to come up with some parameters to characterize a  quantum closed democratic system. But we know of some creative ways to serve specific purposes. The large N expansion is very useful, starting with the Hartree approximation, it produces a pretty good semiclassical theory. The $\epsilon = 4-D$ (D being the dimension of spacetime) expansion worked well in critical phenomena, and dimensional regularization (identifying ultraviolet divergences by targeting terms proportional to powers of 1/(D-4))  was invented to provide a proof of the renormalizability of non-Abelian gauge theories in a covariant way. Here, we wish to study how a hierarchical organization or representation can provide clues to understanding the behavior of a closed democratic system.     

\subsection{Structural versus Representational Hierarchy}

Many physical objects have well-defined or easily identifiable levels of structure. Physicists for centuries have been probing these levels of structure, layer by layer, from molecules to atoms to nucleus to nucleons to quarks and gluons, and more. What at one time was considered as elementary turned out to be composites. With this structural hierarchy it is easy to introduce approximations to focus on the activities of  those particles of interest at a specific level (from nuclear/particle to atomic/molecular physics) and pretty much ignore the complex underlying structures.  Ignore is an exaggeration of course, theoretically speaking,  but in essence this is implemented operationally.  By invoking the Born-Oppenheimer approximation to separate the nuclear motion from the electronic motion one carries a small electron to nucleus mass ratio. Or, to derive a low energy effective field theory, there are terms proportional to the light to heavy mass sector ratio we need to be mindful of \cite{CH97}. Similarly for the gauge hierarchy problems or extra dimensional (Kaluza-Klein) unification theories. Physical systems with an innate structural hierarchy are relatively easy to handle.  

Another kind of hierarchical ordering is more in the nature of representation than structural.  Representation is created to facilitate a better organization of the information about the system which we can extract in an easier way.   Consider the molecular gas. It is a closed democratic system -- all particles are the same, there is no distinguished party in the system and there is no environment. How do we get a handle on it to perform some analysis? We need to introduce some scale of physical relevance. At this point we may immediately hear objections to the effect that any scale one  interjects into the system is subjective, and the description which one gives depends on one's preferred choice in how to represent this system. When the perceptions are  different  how can everyone agree on a unique objective physical reality? 

These are important issues to bear in mind. Indeed this issue is even more acute when we dwell on physical reality from quantum histories which  interfere constantly, or, in quantum measurement, where the measuring device not just bear witness to, but also partakes of the `collapse of the wave function' and shapes the observed reality.  That is why we need to appeal to hydrodynamic variables, they give our classical minds some peace. Putting quantum worries aside for now, let us just focus on a classical closed democratic system. It is reassuring that there are types of representation which have realistic physical meanings and capable of producing verifiable or falsifiable consequences.  Correlation (BBGKY) hierarchy is an example and a good reason why Boltzmann's arduous journey remains inspirational today.


\subsection{Correlation Hierarchy}

At the microscopic level (of molecular dynamics) all molecular movements are time-reversal invariant, not so at the macroscopic level -- dissipation and violation of time-reversal invariance are observed. To reconcile this difference and understand the origin of dissipation Boltzmann came up with the idea that if only one particle distribution functions  observed, and the molecular chaos assumption is imposed for binary collisions, then he can explain the origin of dissipation as a rather common macroscopic phenomena. The BBGKY hierarchy of succesive higher order correlation functions is a suitable way to systemize the information in the gas. It is only upon the truncation of the hierarchy and the imposition of a causal factorization condition -- what we  called `slaving'  (the re-expression of the higher correlation functions in terms of the lower ones, assuming that colliding partners are uncorrelated initially)-- that the closed system represented by the full hierarchy is rendered effectively open, and with it enters dissipation. We shall show that dissipation is accompanied by noise accounting for the contributions of the higher order correlation functions. Thus the name `correlation noise'. 

The reason why we say a representation by the correlation hierarchy produces physically unambiguous results is the following: The `freedom' in choosing which order of correlation one wishes to measure is entirely up to the experimenter. Oftentimes she is not quite free to choose because her observation is limited by the precision of her measuring devices. If her instrument can only measure the one particle distribution function, well, all information about two particle correlations is lost to her.  This is so only for specific processes taking place under specific conditions. What is observed is not subjective, because the same results can be reproduced under the same conditions by anyone. Yes, observers with different degrees of precision will report on different phenomena. If I don't have access to a particle accelerator whether I believe the nucleons are elementary or are composites is totally irrelevant. Subjective views have no place here.  
		
		Now we can describe at least conceptually how decoherence in a closed quantum system is implemented in a hierarchical representation.  Decoherence of the sector of  lower  (up to the nth)  order correlation functions   by the quantum fluctuations associated with the sector of higher  (n+1th and above) order correlation functions can be represented by a correlation noise of the nth order. It is in this sense that a closed democratic system (say of N molecules) can be treated as an effectively open system, with the lower sector acting as the system of interest and the higher order sector its environment.
The effective environment thus changes with the correlation order stipulated by the experimenter or observers at the upper limit of precision in her measurements.  The coarse-graining carried out in a graded environment by correlation order is an example of representational hierarchical coarse-graining. 
				
		In the next section we shall consider another kind of hierarchy which has both structural and representational elements. We will see that not only the way we choose to organize the structural components in our system, but also the order we coarse-grain different components, lead to very different results.  This is a type of structural hierarchical coarse-graining.


\section{\large Hierarchical Coarse-Graining}

We mentioned earlier that almost all macroscopic or mescoscopic objects in nature have hierarchical levels of structure, and as such, theories operating at a particular level are effective theories, results of coarse-graining over the underlyging substructures. How good an {effective theory} is in its depiction of physical phenomena at a particular scale is usually determined by the appropriateness in the choice of the collective variables, the correctness in the choice of coarse-graining for the targeted goal, its extent in relation to the probing scale and the precision of measurement.  We talked about the first (collective variables) and the third elements (measurement precision). Here we shall focus on the second element, coarse-graining, which to me is probably the most subtle yet important factor. 
The issues involved are quite extensive. I listed them in an essay for the 1993 Time-Asymmetry conference \cite{Timeasy} (reproduced partly in Chapter 1 of \cite{CH08}). One key issue concerns \textit{persistent emergent structures} in a closed quantum system: How does the degree of persistence depend on a) the choice of collective variables, b) the CG chosen for each layer of substructure, c) stability with respect to successive CG, and d) robustness with respect to variations of CG measures?  I can only partially address a subset of these issues here. (Similar concerns seem to be shared in \cite{GelHar07}   from the angle of `quasi-classical coarse-graining'.)

When the basic constituents or the components of a system are known it is easier to tackle this problem because the substructures are already persistent.  For concreteness let us take as example the closed quantum system with three basic components: an atom, a quantum field and a dielectric medium. When we say the correctness in the choice of coarse-graining for the targeted goal, we ask, which components shall be coarse-grained, and in what order. That depends on what we want. In the example below our target is the Casimir-Polder force between the atom and the dielectric.  We know the whole atom feels the force but it is only its internal degrees of freedom (the electronic levels) which interact  with the field, in fact, the fluctuations in the quantum field.  Thus we need two variables for the atom: its external or mechanical dof, usually the center of mass (CoM). For the quantum field, we know its vacuum state is changed when there is a medium present, thus we need to know how the medium modifies the field. As for the medium, the dielectric modeled by oscillators on a lattice also has two sets of variables: the polarization vector and the bath variables which damp the motion of the medium oscillators, preventing run-away behavior. The  polarizability is one of the many dynamical susceptibility functions describing the electromagnetic responses of different materials in different capacities.  Ordinarily one may choose a shortcut which lumps the first two steps by starting with a phenomenological response function as is done in Lifshitz's macroscopic electrodynamics \cite{MQED} -- there sits an example of CG at the innermost layer.

To illustrate how different targets demand  different CG, or the ordering of CG, let's leave the medium aside and just consider an atom in a quantum field. So in this case we have only three variables. Suppose we want to see how the motion of an atom affects its spontaneous emission rate --some may consider this effect as \textit{motional decoherence} \cite{ShrHu}. Of these three components, which one should we CG over? The field. The simple reason is because we are asking a question which involves only the two variables of the atom: how does the atom's CoM (edf) motion affect its electronic activities (idf). But more importantly, let's reason out the physics. When an atom moves, it picks up a different field configuration at a different time and space. This changing field configuration is what the atom's internal degrees of freedom respond to, resulting in motional decoherence. 

An  easy  example for this moving atom- quantum field set up: Suppose we wish to calculate the \textit{entanglement} between a moving two-level atom and a quantum field, which variable should be CG over?  The atom's edf.  

Going in the opposite, somewhat more complex, direction is \textit{quantum friction} \cite{qfric}. A neutral atom moving very close to the surface of some dielectric material disturbs the field in the medium. Unlike in the Casimir-Polder force where it is sufficient to consider a stationary medium modifying the field, in quantum friction , one can at best assume the whole system is in a \textit{nonequilibrium} steady state. The field in the dielectric back-reacts in a highly non-Markovian manner on the atom and results in slowing the atom down. It is interesting to consider how CG should be carried out in an ordered manner, in stages, capturing the interplay of all the variables of the participating entities.  


\subsection{An example: Atom-Field-Medium Interaction}

The material in this section is excerpted from the work of Behunin and Hu \cite{BH11} where the Casimir-Polder force is derived from a microscopic model with three entities: the atom, the quantum field and the medium, represented by five micro/meso-variables: two for the atom, two for the medium.   We want to use the procedure laid out there to illustrate how hierarchical coarse-graining plays an essential role in the emergent physics. The ordering of coarse-graining procedures is embodied in the graded influence action. For readers who don't care about the technical details they can skip from the next subsection on, and just have a look at Table I. For those who want more, you can find a similar derivation of the atomic forces in \cite{BH10}. 

The action describing the entire system $S[ \vec{z}, \vec{Q}, A^\mu,
\vec{P}, \vec{X}_\nu ]$ is the sum of eight terms:
(We adopt the Einstein summation convention and natural units throughout $\hbar = c = k_B =1$.)
\begin{equation}
S[ \vec{z}, \vec{Q}, A^\mu, \vec{P}, \vec{X}_\nu] \equiv S_Z + S_Q + S_E + S_M + S_X+  S^{AF}_{int} + S^{PF}_{int} + S^{PX}_{int},
\end{equation}  with five free actions pertaining to the five dynamical variables and three interaction actions. 
Here 1) $S_Z$ is the action for the motion of the atom's center of mass with coordinate  $\vec{z}$ and total mass $M$ under the
influence of an external potential $V[\vec{z}(\lambda)]$:

\begin{equation}
\label{ }
S_Z[\vec{z}]=  \int_{t_i}^{t_f} dt  \ \bigg[  \frac{M}{2} {\dot{\vec{z}}}^2(t)-V[\vec{z}(t)] \bigg]
\end{equation}
where $t$ is the atom's worldline parameter. 2) The internal
degrees of freedom of the atom are modeled by a three dimensional
harmonic oscillator with coordinate $\vec{Q}$ and natural frequency
$\Omega$ with action

\begin{equation}
\label{ }
S_Q[\vec{Q}]= \frac{\mu}{2}  \int_{t_i}^{t_f} dt  \ \left[ \dot{\vec{Q}}^2(t)-\Omega^2 \vec{Q}^2(t) \right] 
\end{equation} where $\mu$ is the oscillator's reduced mass. 3) The dynamics of
the free photon field is described by $S_E$ where `E' stands for the
electromagnetic field, 

\begin{equation}
\label{ }
S_E[A^\mu]= \frac{1}{4}\int d^4 x \  F^{\mu \nu} F_{\mu \nu}.
\end{equation} where $F_{\mu \nu}= \partial_\mu A_\nu-\partial_\nu A_\mu $ is the field strength tensor, with $A_\mu$ the photon field's vector potential and $\int d^4 x = \int_{t_i}^{t_f} dt \int d^3 x$.
4) The matter may be described by  a continuous lattice of harmonic oscillators with natural frequency $\omega$ which is meant to model the polarization of the medium, the coordinate of each oscillator is described by
the vector field, $\vec{P}$,  with action

\begin{equation}
\label{ }
S_M[\vec{P}]= \frac{1}{2}\int_V d^4 x[ \dot{\vec{P}}^2(x)-\omega^2 \vec{P}^2(x)].
\end{equation}
The subscript $V$ on the integration denotes that 
the spatial integration is restricted to the volume containing the matter.

5) Each oscillator comprising the matter field is coupled to a reservoir. The 
reservoir is composed of a collection of oscillators at each point 
with frequency dependent mass $I(\nu)$ and coordinates $\vec{X}_\nu$,
 with natural frequency
$\nu$

\begin{equation}
\label{ }
S_X[\vec{X}_\nu]= \frac{1}{2}\int_V d^4 x \int d \nu \ I(\nu)[ \dot{\vec{X}}_\nu^2(x)-\nu^2 \vec{X}^2(x)].
\end{equation}

The interaction of these parties is specified by the three interaction actions. 
6) The interaction between
the internal degree of freedom (dof) of the atom and the field is

\begin{equation}
\label{ }
S^{AF}_{int}[A^\mu,  \vec{Q}, \vec{z}] = q \int d\lambda \  Q^i(\lambda) E_i(\lambda, \vec{z}(\lambda))
\end{equation}
where $q$ represents the electronic charge. For the case of the
matter the dipole moment of each oscillator is coupled with the local
electric field with a coupling $\Omega_P$

\begin{equation}
\label{ }
S^{PF}_{int}[A^\mu, \vec{P}] = \Omega_P  \int_V d^4 x \ P^i_\nu(x) E_i(x).
\end{equation}
For our model, which considers only the coupling of the 
electric field to the local polarization of the matter, there is no magnetic response
for the medium and so the permeability can assume its vacuum value, $\mu_o$, throughout.
 It should be noted that we have \textit{not}
included interactions among the elements of the dielectric which will provide 
a spatially local form for the permittivity in the macroscopic Maxwell's equations.

Finally, each oscillator composing the matter is coupled to a reservoir with frequency 
dependent `charge' $g(\nu)$ which will provide dissipation and noise

\begin{equation}
\label{ }
S^{PX}_{int}[ \vec{P}, \vec{X}_\nu] = \int_V d^4 x \int_0^\infty d\nu \ g(\nu) P_i (x) X^i_\nu(x).
\end{equation}



We can divide our considerations into two stages: In Stage I we will calculate the effect of the medium on the field. We then make a rest stop, show where the popular Lorentz, Drude and plasma models of the medium arise from, and where the semi-phenomenological `Macroscopic QED' approach of Lifshitz \cite{MQED} resides. In Stage II, we first find out how the medium-modified field interacts with the atom's internal degrees of freedom, and then derive the Casimir-Polder force the atom's center of mass (mechanical or external degree of freedom) experiences. 

\subsection{The medium and the medium-modified field}

The `medium' refers to the combined reservoir + matter system. A reservoir for the network of oscillators modeling the dielectric is needed to damp out their motion and keep them in a stationary state. 
Thus two steps need be taken in Stage I: IA) how the reservoir alters the medium, IB) how the medium alters the quantum field. 

After step IA, namely, after coarse-graining over the reservoir the authors of \cite{BH11} obtain the semiclassical equation of motion for the polarization vector of the matter,
 
\begin{equation}
\label{MEOM}
\ddot{P}^k + \omega^2 P^k - \int_{t_i}^{t_f} dt' \frak{G}^{ret}(t,t') P^k(t') = \xi^k_X(t).
\end{equation}
where $\frak{G}^{ret}$ is the retarded Green function of the reservoir field, and $\vec{\xi}_X$ is the classical stochastic force or noise driving the matter with probability distribution described by $\mathcal{P}[\vec{\xi}_X]$. Due to the Gaussianity of $\mathcal{P}[\vec{\xi}_X]$ all of its moments are specified by the mean $\left< \xi_X^j \right>_{\xi_X}=0$ and the variance 
\be \left<\{ \xi_X^j(x), \xi_X^k(x') \} \right>_{\xi_X}= \delta^{jk} \frak{G}_H(x, x'), \te where  $\frak{G}_H$ is the Hadamard function of the reservoir field, and 
\be
\left< ... \right>_{\xi_X} \equiv 
\int \mathcal{D} \vec{\xi}_X \mathcal{P}[\vec{\xi}_X] (...). 
\te

Thus Behunin and Hu show how coarse-graining over the medium degrees of freedom leads to a permittivity and a classical stochastic source responsible for an additional induced component of field fluctuations. 
For the particular case of local fluctuations they found that the stochastic semi-classical action for the field 
takes the same form as the noninteracting field (with frequency dependent velocity) driven 
by an external current (stochastic force). 
(A detailed description of the field's microstate is unimportant in the
description of the atom-surface force.) 
From the action we obtain  the stochastic semi-classical equations of motion for the medium-modified field 

\begin{equation}
\label{SCEM} \nabla \times \nabla \times \vec{A}({\omega'}, \vec{x}) -
{\omega'}^2(1+ \Omega_P^2 \tilde{g}_{ret}({\omega'}, \vec{x})) \vec{A}({\omega'}, \vec{x}) =  i \Omega_P {\omega'}
 \tilde{g}_{ret} ({\omega'}, \vec{x}) \vec{\xi}_X({\omega'},\vec{x}) .
\end{equation}
where $\tilde{g}_{ret}$ is the (Fourier-transformed) retarded Green function of the medium and $\vec{\xi}_X(x)$ is the stochastic force from the fluctuations of the medium.  By comparing with the macroscopic Maxwell's equations one can readily identify the
permittivity  as $\varepsilon({\omega'},
\vec{x})=1+\Omega_P^2 \tilde{g}_{ret}({\omega'}, \vec{x})$ 
($\vec{\xi}_M(x)$ drives the field in addition to the reservoir when the system is not in a steady-state).
 
We can take a rest stop now at the end of Stage I and take stock of what our micro-physics model delivers:   For Ohmic spectral density of the reservoir the frequency-dependent permittivity corresponds with the \textit{Lorentz oscillator model}
\begin{equation} \label{Lorentz}
\varepsilon({\omega'}, \vec{x}) = 1 - \frac{\Omega_P^2}{\omega' (\omega' + i \gamma) - \tilde{\omega}^2} \ \ \ \ \  \vec{x} \in V. \end{equation}
When the restoring force of the matter vanishes we obtain the \textit{Drude model} and the \textit{plasma model} when $\gamma \rightarrow 0$  in addition. In this form it's clear that the matter-field coupling  $\Omega_P$ can be interpreted as the medium's plasma frequency. 

One can see from equation (\ref{SCEM}) a striking similarity with \textit{Lifshitz's Macroscopic QED}. Due to the linearity of our theory this is not surprising. Indeed if one were to proceed from this point treating the field semi-classically and choosing all space to be filled with a dielectric material, albeit in vacuum this dielectric is fictitious, one would exactly reproduce the predictions of MQED using (\ref{SCEM}) and choosing the dielectric to be in a thermal state\footnote{However, it is important to note that in this case the stochastic field $\xi^j_X$ represents the fluctuations of the medium only and does not include the intrinsic fluctuations of the field. After the next level of coarse-graining described below the intrinsic quantum fluctuations of the field will enter which is different from the induced fluctuations from interaction with the dielectric medium.}.

\subsection{Medium-modified quantum field's effect on the atom}

For the second stage, the task is to find out how this medium modified field affects the atom.  This also involves two steps: IIA) how the field interacts with the atom's internal degrees of freedom (here using an oscillator model), IIB) calculate the force exerted on the atom  (its mechanical or external degree of freedom) by coarse-graining over the atoms' internal degrees of freedom.  

\subsubsection{Medium-modified quantum field}

By coarse-graining over the medium-influenced field we can incorporate the averaged effect of the field on the atom's trajectory without a specific knowledge of the final field state, leading to the reduced density matrix for the atom
\begin{eqnarray}
\label{ }
&&\rho_r(z_f,z_f', \vec{Q}_f, \vec{Q}_f'; t_f)= 
 \int^{\vec{z}_f, \vec{z}_f'}_{CTP} \mathcal{D} \vec{z}  
 \int^{\vec{Q}_f, \vec{Q}_f'}_{CTP} \mathcal{D} \vec{Q}  
  e^{\{ i (S_Z[\vec{z}]  -S_Z[\vec{z}'] +S_Q[\vec{Q}] -S_Q[\vec{Q}'] \}}
    \nonumber \\  
 &&   \times
 \int  \mathcal{D} A^\mu_f  
  \int^{A^\mu_f, A^{\mu}_f}_{CTP} \mathcal{D}A^\mu \
e^{\{ S^{AF}_{int}[ \vec{z}, \vec{Q},  A^\mu]-S^{AF}_{int}[ \vec{z}' , \vec{Q}' ,  A^{\mu'}] \}} \mathcal{F}_M[A^\mu, A^{\mu'}]
\end{eqnarray}
(CTP denotes that it is a closed-time-path \cite{ctp} integral) where the first integral in the second line traces over the final field configurations.

For the assumed initially factorized state the path integrals
over the field can be evaluated exactly if the initial state is
Gaussian yielding the field-reduced influence functional,
\begin{equation}
 \mathcal{F}_\xi[J^\mu, J^{\mu'}, \xi] = \int \mathcal{D} \vec{\xi}_X \mathcal{P}[\vec{\xi}_X]  \int \mathcal{D} \vec{\xi}_M \mathcal{P}[\vec{\xi}_M] \exp\{ iS^E_{IF}[J^\mu, J^{\mu'},
\xi] \} 
\end{equation}
 expressed in terms of the  influence action, $S^E_{IF}$, given by
\begin{eqnarray}
\label{SEIF}
&& S^E_{IF}[J^\mu, J^{\mu'}, \xi^j]=
 \int d^4 x \int d^4 x' J^{\mu-}({x}) \nonumber \\  
&& \times \bigg[ \tilde{D}^{ret}_{\mu \nu}({x}, {x}') \big(J^{\nu+}({x}')-\kappa^\nu_i \xi^i({x}')\big)
  +\frac{i}{4} \tilde{D}^{H}_{\mu \nu}({x},{x}') J^{\nu-}(\vec{x}') \bigg]
\end{eqnarray}
where $\vec{\xi}(x)  \equiv     \vec{\xi}_M(x) + \int_{t_i}^{t_f} dt' \ \tilde{g}_{ret} (t,t') \vec{\xi}_X(t',\vec{x}) $.
%
%
The current density $J^\mu$ in (\ref{SEIF}) comes from the atom-field interaction and takes the explicit form
\begin{equation}
\label{ }
J^\mu(x)=-q \int d\lambda \ Q^i(\lambda) \kappa_i^\mu \delta^4(x^\alpha-z^\alpha(\lambda))
\end{equation}
where the derivative operator $\kappa^\mu_i=-\partial_0 \eta^\mu_i+\partial_i \eta^\mu_0$ yields the electric field when contracted with the vector potential $E^i =\kappa_\mu^i A^\mu$ and also enforces current conservation $\partial_\mu \kappa^\mu_i f(x)=(-\partial_0 \partial_i+\partial_i \partial_0)f(x)=0 $.
The integral kernels $\tilde{D}^{ret}_{\mu \nu}$ and $\tilde{D}^{H}_{\mu \nu}$ are
the retarded  Green's and Hadamard function for the medium-altered
electromagnetic field which result from solving the semi-classical
equations of motion (for the retarded Green's function sourced by a delta function). 
The retarded Green's function describes the classical electrodynamical propagation
of the field in the presence of the dielectric material, and the Hadamard function describes
the field's intrinsic quantum fluctuations.

In the temporal gauge the semi-classical equation of motion for the field's retarded Green's
function  in the presence of a dielectric medium satisfies 
\begin{eqnarray}
\label{GFeqn}
&& \epsilon^{ab}_{\ \ \ i}   \epsilon^{mn}_{\ \ \
\ b} \partial_a \partial_m  \tilde{D}^{ret}_{nk}(x,x')  + \frac{\partial^2}{ \partial t^2 }   \tilde{D}^{ret}_{ik}({x}, {x}')  \nonumber \\  
&+&  \frac{\partial}{ \partial t} \int_{t_{in}}^{t_f} dt_2 \ \tilde{g}_{ret}(t,t_2;\vec{x})
\frac{\partial}{ \partial t_2}
\tilde{D}^{ret}_{ik}(t_2 ,\vec{x},  x') =  \delta_{ik} \delta^4({x}-{x}')
\end{eqnarray}
%
%
where $\epsilon_{abc}$ is the Levi-Civita symbol (Roman indices
refer to spatial components).
The solution to (\ref{GFeqn}) gives the particular solution, $A^P_j$, to the semiclassical equation of motion (\ref{SCEM}) for the electromagnetic field
\begin{equation}
\label{ }
A^P_j(x)= -   \int d^4 x'   \tilde{D}^{ret}_{jk}(x, x') \frac{ \partial }{ \partial t'}  \xi^k( x').
\end{equation}

As was noted previously the Heisenberg equations of motion for the field operators take the same form as the semi-classical equation of motion because of the linear coupling

\begin{equation}
\label{HEOM}
 \epsilon^{ab}_{\ \ \ i}   \epsilon^{mn}_{\ \ \
\ b} \partial_a \partial_m  \hat{A}_n(x)  + \frac{\partial^2}{ \partial t^2 }   \hat{A}_i(x)  +  \frac{\partial}{ \partial t} \int_{t_{in}}^{t_f} dt_2 \ \tilde{g}_{ret}(t,t_2;\vec{x})
\frac{\partial}{ \partial t_2}
 \hat{A}_i(t_2, \vec{x})   =  \frac{ \partial }{ \partial t}  \xi_i( x)
 \end{equation}
 therefore we can use the homogeneous solution to (\ref{HEOM}) to construct the symmetric two-point function of the field operators to give the Hadamard function 
 \begin{equation}
\label{ }
 \tilde{D}_{H}^{i j}(x, x') = \left<  \{ \hat{A}^i_o(x),  \hat{A}^j_o(x') \} \right>. 
\end{equation}

Finally, the field-reduced density matrix takes the form
\begin{eqnarray}
\label{ }
\rho_r(z_f,z_f', \vec{Q}_f, \vec{Q}_f'; t_f)=  
 \int^{\vec{z}_f, \vec{z}_f'}_{CTP} \mathcal{D} \vec{z}  
 \int^{\vec{Q}_f, \vec{Q}_f'}_{CTP} \mathcal{D} \vec{Q} 
 & e^{i (S_Z[\vec{z}] +S_Q[\vec{Q}]-S_Z[\vec{z}'] -S_Q[\vec{Q}'])} \nn \\ & \times \mathcal{F}_\xi[J^\mu, J^{\mu'}, \xi].
\end{eqnarray}

\subsubsection{C-P force by coarse-graining the atom's internal dof}

At this point we have the density matrix that describes the dynamics
of the atom's trajectory and its internal degrees of freedom under
the influence of the medium-altered field. As with the previous parties it is only the
averaged effect and not the microscopic details of the atom's
internal degree of freedom which we need for the description of the
force. 

In our last tier of coarse-graining we
trace over $\vec{Q}_f$ the oscillator's internal dof to obtain the
oscillator-reduced density matrix which characterizes the
dynamics of the atom's trajectory determined by its interaction with all remaining parties
\begin{equation}
\label{rhoZ}
\rho_Z(z_f,z_f'; t_f)=  
 \int^{\vec{z}_f, \vec{z}_f'}_{CTP} \mathcal{D} \vec{z} \ e^{i (S_Z[\vec{z}] -S_Z[\vec{z}'] )} \mathcal{F}_Z[\vec{z}, {\vec{z}}'].
\end{equation}
The environmental influences are now packaged in the
oscillator-reduced influence functional $\mathcal{F}_Z[\vec{z}, \vec{z}']$ with their back-action accounted for in a self-consistent manner. The coarse-grainings are summarized in Table \ref{table:CG}
\begin{table}[ht] 
\caption{Layers of Coarse-Graining and Physics Remaining at Each Tier} 
\centering 
\begin{tabular}{ccccc} 
\hline\hline 
\ Tier  &  C-Grain  &  Infunc'l \ \ \ & \ \ \ Detailed Physics \ \ \  & \ \ \ Remainder \ \  \\ [1.5ex] 
0 &                &                                 & \text{atom 2 + field + matter 2}                     & $\vec{z}, \vec{Q}, A^\mu, \vec{P}, \vec{X}_\nu$    \\
1 & $Tr_X$ & $\mathcal{F}_X$   & \text{atom + field + polarization}  & $\vec{z}, \vec{Q}, A^\mu, \vec{P}$    \\
2 & $Tr_P$ & $\mathcal{F}_M$  & \text{atom + field}                  & $\vec{z}, \vec{Q}, A^\mu$  \\
3 & $Tr_A$ & $\mathcal{F}_\xi$ & \text{atom's ext \& int dof}                              & $\vec{z}, \vec{Q}$  \\ 
4 & $Tr_Q$ & $\mathcal{F}_Z$  & \text{atom's motion}              & $\vec{z}$ \\ [1ex]
\hline 
\end{tabular} 
\label{table:CG} 
\end{table}
at each tier the influence of all lower tiers on the remaining degrees of freedom is packaged in the influence functional. 

We proceed by evaluating $\mathcal{F}_Z[\vec{z}, \vec{z}']$ perturbatively to lowest order in the atom-field coupling. In the fully dynamical case an exact calculation is not possible because the back-action of the field on the dynamics of the atom's dipole moment enters as a third time derivative and includes multiple reflections between the dielectric medium and the atom. 
At leading order in powers of the atom-field coupling, an expression for the force can be found which neglects the radiation reaction to the atom's dipole moment and the effects of multiple reflections.

To ease the notational burden we define
\begin{equation}
\label{ }
\left<...\right>_o = \int_{-\infty}^\infty d\vec{Q}_f  
 \int^{\vec{Q}_f, \vec{Q}_f}_{CTP} \mathcal{D} \vec{Q} 
  \ e^{i (S_Q[\vec{Q}]-S_Q[\vec{Q}'])} (...)
\end{equation}
which represents the noninteracting time-dependent expectation
value with respect to the oscillator's initial state. With this
simplification the oscillator-reduced influence functional can be
compactly expressed in terms of quantum and stochastic expectation
values
\begin{equation}
\label{IFZ}
e^{iS_{IF}[\vec{z}, \vec{z}']}  \stackrel{def}{=}  \mathcal{F}_Z[\vec{z}, \vec{z}']= \bigg< \bigg<  \bigg< e^{i S^E_{IF}} \bigg>_o \bigg>_{\xi_M} \bigg>_{\xi_X}
\end{equation}
which introduces the influence action, $S_{IF}$. With it we get the coarse-grained effective action, $S_{CGEA}[\vec{z},\vec{z}'] \equiv
S_{Z}[\vec{z}]-S_{Z}[\vec{z}']+S_{IF}[\vec{z},\vec{z}']$,

At the saddle point of (\ref{rhoZ}) we get the semi-classical equation of motion for the atom's trajectory: 
\begin{equation}
\label{ }
\frac{\delta}{ \delta z^{k-}(\tau)} S_{CGEA}[\vec{z},\vec{z}']\bigg|_{z^{k-}=0}=0 \Rightarrow M\ddot{z}_k(\tau)+\partial_k V[\vec{z}(\tau) ]=f_k(\tau),
\end{equation}
whereas the influence force, $f_k(\tau)$, is given by
\begin{equation}
\label{ }
\frac{\delta}{ \delta z^{k-}(\tau)} S_{IF}[\vec{z},\vec{z}']\bigg|_{z^{k-}=0}=f_k(\tau),
\end{equation}
which is the sought-after Casimir-Polder force.

\section{\large Quantum Hierarchical System: Correlation Noise}



In this section we implement what was described in Sec. III, aiming at deriving the correlation noise from the higher order correlation functions in the Schwinger-Dyson hierarchy of an interacting quantum field.  We first summarize the conceptual framework schematically, drawing the classical parallel of stochastic Boltzmann equation \cite{StoBol}.  We then define our goal and present the methodology to reach that goal.  Two subtopics used or related here are: The application of 2PI effective actions to quantum kinetic field theory \cite{CH88}, which is the starting point toward the construction of a master effective action \cite{CH95}. For decoherent correlation histories, see \cite{CH93,Ana97}. Further details of the subject matter covered here are in \cite{CH99,Cal09}. (See also related work in, e.g., \cite{Reinhard,Greiner}.)

Finding an effective stochastic formulation to describe an interacting quantum system has a much broader scope and significance. Stochastic inflation \cite{Starob} and stochastic gravity \cite{StoGra} are two noticeable themes in gravity and cosmology.

\paragraph{Conceptual Framework}

1) Classical:  Boltzmann equation is the lowest order BBGKY hierarchy.  

2) BBGKY hierarchy governing the correlation functions $\Rightarrow$  Schwinger - Dyson hierarchy of equations (SDE) of interacting quantum field theory. 

3)  Higher order correlation functions represented as a noise -- the correlation noise.  This is the key link. 

4) The inclusion of this noise is necessary by demand of the fluctuation-dissipation relation. This gives rise to a stochastic Schwinger-Dyson hierarchy,  a QFT generalization of the stochastic BBGKY hierarchy, the lowest order being the stochastic Boltzmann equation. 


Explanation:  Points 1) \& 2) As described earlier, in classical kinetic theory the full BBGKY hierarchy gives complete information of the closed system, say, of a molecular gas. It is upon A) the truncation of the hierarchy and B) the imposition of causal factorization conditions -- the combination of these two procedures we call ''slaving'', the molecular chaos assumption being a familiar example -- that the equations for the low order correlation functions (such as the Boltzmann equation for the one particle distribution function) acquires dissipative behavior. 

Points 3) \& 4). While the lower- order correlation functions constitute the system of interest, the slaved higher order correlation functions play  the role of an environment of this effectively open system, their fluctuations being the source of noise in the stochastic Boltzmann equations.
 This stochastic generalization of the Boltzmann equation gives rise to a Boltzmann-Langevin equation and its field theoretical parallel is the stochastic Schwinger-Dyson equation.

\paragraph{Our goal}  is to obtain a self-contained dynamics for the propagators. What this means is that the propagators or correlation functions should carry in them the effect of their interaction with the higher correlation functions as embodied in the BBGKY or Dyson-Schwinger hierarchy. The procedure which makes this possible is slaving. (For Boltzmann, the imposition of the molecular chaos assumption). The necessary consequence is the appearance of dissipative behavior in the dynamics of the correlation functions, and, owing to the existence of a dissipation-fluctuation relation, the necessary existence of noise as well.

\paragraph{Pathway}

We start with classical kinetic theory and show a derivation of the stochastic Boltzmann equation. In the next subsection we shall sketch the main steps towards deriving a Boltzmann - Langevin for interacting QFT. The key link is correlation noise. The final stop is decoherence brought about by this noise, which results from the coarse-graining of the higher correlation functions.  Again, for readers wary of too much technical details, here is a useful CG scheme for the rest of this section: After a glance of the classical stochastic Boltzmann Eqs. (\ref{lanbe}) with (\ref{singucor}) and (\ref{sigmasq}),  go to Sec. 5.2 for the quantum field equivalent. All the formal essentials are there.

\subsection{Boltzmann equation for a classical relativistic gas}

Consider a dilute gas of relativistic classical 
particles. The system is described by its one particle
distribution function $f\left( X,k\right) $, where $X$ is a position
variable, and $k$ is a momentum variable. Momentum is assumed to lie on a
mass shell $k^2+M^2=0$. (We use the MTW convention, with signature -+++ for
the background metric) and have positive energy $k^0>0$. In other
words, given a spatial element $d\Sigma ^\mu =n^\mu d\Sigma $ and a momentum
space element $d^4k$, the number of particles with momentum $k$ lying within
that phase space volume element is

\begin{equation}
dn=-4\pi f\left( X,k\right) \theta \left( k^0\right) \delta \left(
k^2+M^2\right) k^\mu n_\mu \;d\Sigma \frac{d^4k}{\left( 2\pi \right) ^4}
\label{distrif}
\end{equation}

The dynamics of the distribution function is given by the Boltzmann equation,  in a notation adapted to our later needs:   
\begin{equation}
k^\mu \frac \partial {\partial X^\mu }f\left( k\right) =I_{col}\left(
X,k\right)  \label{be}
\end{equation}

\begin{equation}
I_{col}=\frac{\lambda ^2}4\left( 2\pi \right) ^3\int \left[ \prod_{i=1}^3%
\frac{d^4p_i}{\left( 2\pi \right) ^4}\theta \left( p_i^0\right) \delta
\left( p_i^2+M^2\right) \right] \left[ \left( 2\pi \right) ^4\delta \left(
p_1+p_2-p_3-k\right) \right] {\bf I}  \label{icol}
\end{equation}

\begin{equation}
{\bf I}=\left\{ \left[ 1+f\left( p_3\right) \right] \left[ 1+f\left(
k\right) \right] f\left( p_1\right) f\left( p_2\right) -\left[ 1+f\left(
p_1\right) \right] \left[ 1+f\left( p_2\right) \right] f\left( p_3\right)
f\left( k\right) \right\}  \label{integ}
\end{equation}

The entropy flux is given by
\begin{equation}
S^\mu \left( X\right) =4\pi \int \frac{d^4p}{\left( 2\pi \right) ^4}\theta
\left( p^0\right) \delta \left( p^2+M^2\right) p^\mu \left\{ \left[
1+f\left( p\right) \right] \ln \left[ 1+f\left( p\right) \right] -f\left(
p\right) \ln f\left( p\right) \right\}  \label{entroflux}
\end{equation}
while the entropy itself $S$ is (minus) the integral of the flux over a Cauchy surface. Now consider a small deviation from the equilibrium
distribution
\begin{equation}
f=f_{eq}+\delta f  \label{dev}
\end{equation}

\begin{equation}
f_{eq}=\frac 1{e^{\beta p^0}-1}  \label{feq}
\end{equation}
corresponding to the same particle and energy fluxes

\begin{equation}
\int \frac{d^4p}{\left( 2\pi \right) ^4}\theta \left( p^0\right) \delta
\left( p^2+M^2\right) p^\mu \delta f\left( p\right) =0  \label{partflux}
\end{equation}

\begin{equation}
\int \frac{d^4p}{\left( 2\pi \right) ^4}\theta \left( p^0\right) \delta
\left( p^2+M^2\right) p^\mu p^0\delta f\left( p\right) =0  \label{enerflux}
\end{equation}
Then the variation in entropy becomes
\begin{equation}
\delta S=-2\pi \int d^3X\int \frac{d^4p}{\left( 2\pi \right) ^4}\theta
\left( p^0\right) \delta \left( p^2+M^2\right) p^0\frac 1{\left[
1+f_{eq}\left( p\right) \right] f_{eq}\left( p\right) }\left( \delta
f\right) ^2  \label{deltaSa}
\end{equation}

In the classical theory, the distribution function is concentrated on the
positive frequency mass shell. Therefore, it is convenient to label momenta
just by its spatial components $\vec p,$ the temporal component being
necessarily $\omega _p=\sqrt{M^2+\vec p^2}>0.$ In the same way, it is
simplest to regard the distribution function as a function of the three
momentum $\vec p$ alone, according to the rule

\begin{equation}
f^{(3)}\left( X,\vec p\right) =f\left[ X,\left( \omega _p,\vec p\right)
\right]  \label{newf}
\end{equation}
where $f$ represents the distribution function as a function on four
dimensional momentum space, and $f^{\left( 3\right) }$ its restriction to
three dimensional mass shell. With this understood, we shall henceforth drop
the superscript, using the same symbol $f$ for both functions, since only
the distribution function on mass shell enters into our discussion. The
variation of the entropy now reads

\begin{equation}
\delta S=-\frac 12\int d^3X\int \frac{d^3p}{\left( 2\pi \right) ^3}\frac 1{%
\left[ 1+f_{eq}\left( p\right) \right] f_{eq}\left( p\right) }\left( \delta
f\right) ^2.  \label{deltaS}
\end{equation}
From Einstein's formula, we conclude that, in equilibrium, the distribution
function is subject to Gaussian fluctuations, with equal time mean square value

\begin{equation}
\left\langle \delta f\left( t,\vec X,\vec p\right) \delta f\left( t,\vec Y,%
\vec q\right) \right\rangle =\left( 2\pi \right) ^3\delta \left( \vec X-\vec 
Y\right) \delta \left( \vec p-\vec q\right) \left[ 1+f_{eq}\left( p\right)
\right] f_{eq}\left( p\right)  \label{classcor}
\end{equation}
Note that this fluctuation formula is quite independent of the processes which
sustain equilibrium; in particular, it holds equally for a free and an interacting gas, since it contains no coupling constants.

\subsubsection{Stochastic source to sustain  fluctuations}

In the interacting case, however, a stochastic source is necessary to
sustain these fluctuations. Following the discussion of the FDR above, we
compute these sources by writing the dissipative part of the equations of
motion in terms of the thermodynamic forces

\begin{equation}
F\left( X,\vec p\right) =\frac 1{\left[ 1+f_{eq}\left( p\right) \right]
f_{eq}\left( p\right) }\frac{\delta f\left( X,\vec p\right) }{\left( 2\pi
\right) ^3}  \label{thermoforce}
\end{equation}
To obtain an equation of motion for $f\left( X,\vec p\right) $ multiply both
sides of the Boltzmann equation Eq. (\ref{be}) by $\theta \left( k^0\right)
\delta \left( k^2+M^2\right) $ and integrate over $k^0$ to get

\begin{equation}
\frac{\partial f}{\partial t}+\frac{\vec k}{\omega _k}\vec \nabla f=\frac 1{%
\omega _k}I_{col}  \label{newbolt}
\end{equation}
Upon variation we get

\begin{equation}
\frac{\partial (\delta f)}{\partial t}+\frac{\vec k}{\omega _k}\vec \nabla
(\delta f)=\frac 1{\omega _k}\delta I_{col}  \label{linearb}
\end{equation}

When we write $\delta I_{col}$ in terms of the thermodynamic forces, we find
local terms proportional to $F\left( k\right) $ as well as nonlocal terms
where $F$ is evaluated separately. We shall keep only the former, as it is
usually done in deriving the ``collision time approximation'' to the
Boltzmann equation (related to the Krook-Bhatnager-Gross kinetic equation), thus we write

\begin{equation}
\delta I_{col}\left( k\right) \sim -\omega _k\nu ^2(X,\vec k)F(X,\vec k)
\label{deltaicol}
\end{equation}
where

\begin{equation}
\nu ^2(X,\vec k)=\frac{\lambda ^2}{4\omega _k}\left( 2\pi \right) ^6\int
\left[ \prod_{i=1}^3\frac{d^4p_i}{\left( 2\pi \right) ^4}\theta \left(
p_i^0\right) \delta \left( p_i^2-M^2\right) \right] \left[ \left( 2\pi
\right) ^4\delta \left( p_1+p_2-p_3-k\right) \right] I_{+}  \label{sigmasq}
\end{equation}
$k^0=\omega _k$, and

\begin{equation}
I_{+}=\left[ 1+f_{eq}\left( p_1\right) \right] \left[ 1+f_{eq}\left(
p_2\right) \right] f_{eq}\left( p_3\right) f_{eq}\left( k\right)
\label{iplus}
\end{equation}

This linearized form of the Boltzmann equation provides a quick estimate of the relevant relaxation time. Let us assume the high
temperature limit, where $f\sim T/M$, and the integrals in Eq. (\ref{sigmasq}%
) are restricted to the range $p\leq M.$ Then simple dimensional analysis
yields the estimate $\tau \sim M/\lambda ^2T^2$ for the relaxation time
appropriate to long wavelength modes .

In \cite{CH99} it is explained why the fluctuation-dissipation theorem (FDT) demands that a stochastic source $j$ be present in the
Boltzmann equation Eq. (\ref{be}) (and its linearized form, Eq. (\ref {linearb})) which should assume the Langevin form:

\begin{equation}
\frac{\partial f}{\partial t}+\frac{\vec k}{\omega _k}\vec \nabla f=\frac 1{%
\omega _k}I_{col}+j(X,\vec k)  \label{lanbe}
\end{equation}
Then

\begin{equation}
\left\langle j\left( X,\vec p\right) j\left( Y,\vec q\right) \right\rangle
=-\left\{ \frac 1{\omega _p}\frac{\delta I_{col}\left( X,\vec p\right) }{%
\delta F\left( Y,\vec q\right) }+\frac 1{\omega _q}\frac{\delta
I_{col}\left( Y,\vec q\right) }{\delta F\left( X,\vec p\right) }\right\}
\label{fdtbe}
\end{equation}

From Eqs. (\ref{deltaicol}), (\ref{sigmasq}) and (\ref{iplus}) we find the
noise auto-correlation 
\begin{equation}
\left\langle j(X,\vec k)j\left( Y,\vec p\right) \right\rangle =2\delta
^{\left( 4\right) }\left( X-Y\right) \delta \left( \vec k-\vec p\right) \nu
^2(X,\vec k)  \label{singucor}
\end{equation}
where $\nu ^2$ is given in Eq. (\ref{sigmasq}). Eqs. (\ref{singucor}) and (%
\ref{sigmasq}) are the solution to our problem, that is, they describe the
fluctuations in the Boltzmann equation, required by consistency with the
FDT. Observe that, unlike Eq. (\ref{classcor}), the mean square value of the
stochastic force vanishes for a free gas.

In this discussion, of course, we accepted the Boltzmann equation as given without tracing its origin. We now want to see how the noises in Eq. (\ref{singucor}) originate from a deeper level, that related to the higher correlation functions, which we call the correlation noises. We now turn to the statistical mechanics of interacting quantum fields.

\subsection{Schwinger-Dyson Hierarchy for Interacting Quantum Fields}

Consider a scalar field theory $\phi ^a$ (we use DeWitt's condensed notation where the index $a$ denotes both a space - time point and one or the other branch of the closed-time-path (CTP) \cite{ctp}) whose action is 
$S=S\left[ \phi ^1\right] -S^{*}\left[ \phi ^2\right] $. We shall use the two particle irreducible (2PI) representation \cite{CH88,RH97} where  the (two-point) correlation function stands as an independent variable apart from the mean field. Thus there will be an additional separate source $K_{ab}$ driving $\phi ^a\phi ^b$ over the usual $J_a \phi^a$ term in the 1PI representation. Our strategy is to seek a description of the field in terms of a new object  ${\bf G}^{ab}$, namely, a stochastic correlation function whose expectation value over the noise average  gives the usual two point functions reduces to $%
G^{ab},$ while its fluctuations reproduce the quantum fluctuations in the binary products of field operators. 
From the generating functional
\begin{equation}
Z\left[ K_{ab}\right] =e^{iW\left[ K_{ab}\right] }=\int D\phi
^a\;e^{i\left\{ S+\frac 12K_{ab}\phi ^a\phi ^b\right\} }  \label{genfun}
\end{equation}
we have

\begin{equation}
G^{ab}=\left\langle \phi ^a\phi ^b\right\rangle =2\left. \frac{\delta W}{%
\delta K_{ab}}\right| _{K=0}  \label{greenfun}
\end{equation}
and

\begin{equation}
\left. \frac{\delta ^2W}{\delta K_{ab}\delta K_{cd}}\right| _{K=0}=\frac i4%
\left\{ \left\langle \phi ^a\phi ^b\phi ^c\phi ^d\right\rangle -\left\langle
\phi ^a\phi ^b\right\rangle \left\langle \phi ^c\phi ^d\right\rangle \right\}
\label{fluc}
\end{equation}
This suggests viewing the stochastic kernel ${\bf G}^{ab}$ as a Gaussian
process defined (formally) by the relationships

\begin{equation}
\left\langle {\bf G}^{ab}\right\rangle =\left\langle \phi ^a\phi
^b\right\rangle ;\qquad \left\langle {\bf G}^{ab}{\bf G}^{cd}\right\rangle
=\left\langle \phi ^a\phi ^b\phi ^c\phi ^d\right\rangle  \label{stocg}
\end{equation}
Or else, calling

\begin{equation}
{\bf G}^{ab}=G^{ab}+\Delta ^{ab}  \label{delta}
\end{equation}
where
\begin{equation}
\left\langle \Delta ^{ab}\right\rangle =0;\qquad \left\langle \Delta
^{ab}\Delta ^{cd}\right\rangle =-4i\left. \frac{\delta ^2W}{\delta
K_{ab}\delta K_{cd}}\right| _{K=0}.  \label{delta2}
\end{equation}

\subsubsection{Correlation Dynamics from 2PI Effective Action} 

We can define the process $\Delta ^{ab}$ also in terms of a stochastic equation of motion. The Legendre transform of $W$  is the two-particle-irreducible (2PI) effective action (EA)

\begin{equation}
\Gamma \left[ G^{ab}\right] =W\left[ K_{ab}^{*}\right] -\frac 12K_{ab}^{*}%
{\bf G}^{ab};\qquad K_{ab}^{*}=-2\frac{\delta \Gamma }{\delta {\bf G}^{ab}}
\label{tpiea}
\end{equation}
We have the identities

\begin{equation}
\frac{\delta \Gamma }{\delta G^{ab}}=0;\qquad \frac{\delta ^2W}{\delta
K_{ab}\delta K_{cd}}=\frac{-1}4\left[ \frac{\delta ^2\Gamma }{\delta {\bf G}%
^{ab}\delta {\bf G}^{cd}}\right] ^{-1}  \label{ident}
\end{equation}
the first of which is just the Schwinger - Dyson equation for the
propagators; we therefore propose the following equations of motion for $%
{\bf G}^{ab}$

\begin{equation}
\frac{\delta \Gamma }{\delta {\bf G}^{ab}}=-\frac{1}2\kappa _{ab}
\label{lan2pi}
\end{equation}
where $\kappa _{ab}$ is a stochastic nonlocal Gaussian source defined by

\begin{equation}
\left\langle \kappa _{ab}\right\rangle =0;\qquad \left\langle \kappa
_{ab}\kappa _{cd}\right\rangle =4i\left[ \frac{\delta ^2\Gamma }{\delta
G^{ab}\delta G^{cd}}\right] ^{\dagger }  \label{noisecor}
\end{equation}
If we linearize Eq. (\ref{lan2pi}) around $G^{ab}$, then the correlation Eq. (\ref{noisecor}) for $\kappa_{ab} $ implies Eq. (\ref{delta2}) for $\Delta^{ab} $.
Consistent with our recipe of handling $G^{ab}$ as if it were real we should treat $\kappa_{ab}$ also as if it were a real source.

It is well known that the noiseless Eq. (\ref{lan2pi}) can be used as a
basis for the derivation of transport equations in the near equilibrium
limit. Indeed, for a $\lambda \phi ^4$ type theory, the resulting equation is simply the Boltzmann equation for a distribution function $f$ defined from the Wigner transform of $G^{ab}$. It can be shown \cite{CH99} that the full stochastic equation (\ref{lan2pi}) leads, in the same limit, to a Boltzmann - Langevin equation, thus providing the microscopic basis for this equation in a manifestly relativistic quantum field theory
\footnote{Constructing a stochastic representation of quantum field theory has caught increasingly attention. Here's some noteworthy technical points about stochastic equations for the physical propagators, as exemplified in \cite{Cal09}, which nicely completes the quest in \cite{CH99}.  In the 1PI theory, the $\phi^-$ field vanishes identically on-shell. However, in the stochastic approach we assign a nontrivial source $\xi_-^{1PI}$ to it. It is by eliminating this auxiliary field that we recover the usual approach, with a single stochastic source $\tilde \xi$ whose
self-correlation is given by the noise kernel.
Similarly, in the quantum field theory problem the correlator $G^{--} = \langle \varphi^- \varphi^-\rangle$  vanishes identically, as a result of path ordering. However, in the stochastic approach, we consider it as an auxiliary field and couple a source to it. The authors of \cite{CH99} failed to recognize the violation of the constraint $G^{--} = 0$, but guessed the right form for the noise self-correlation by introducing a sign change in their expression arising from adding a hermitian conjugation to the second derivative of the 2PI-EA. This was shown to be unnecessary by Calzetta \cite{Cal09} who provided a satisfactory explanation for this sign change: It is due to the elimination of the auxiliary field $G^{--}$, keeping only the physical degrees of freedom.}.

\subsubsection{Dissipative and Stochastic Dynamics}

Let us first examine some consequences of Eq. (\ref{noisecor}). For a free field theory, we can compute the 2PI EA explicitly 

\begin{equation}
\Gamma \left[ G^{ab}\right] =\frac{-i}2\ln \left[ Det{\bf G}\right] -\frac 12%
c_{ab}\left( -\Box +m^2\right) {\bf G}^{ab}\left( x,x\right)  \label{free2pi}
\end{equation}
where $c_{ab}$ is the $2 \times 2$ CTP metric tensor (see, e.g., \cite{RH97}). We immediately find

\begin{equation}
\frac{\delta ^2\Gamma }{\delta G^{ab}\delta G^{cd}}=\frac i2\left(
G^{-1}\right) _{ac}\left( G^{-1}\right) _{db}  \label{secondvar}
\end{equation}
therefore

\begin{equation}
\left\langle \Delta ^{ab}\Delta ^{cd}\right\rangle =i\left[ \frac{\delta
^2\Gamma }{\delta G^{ab}\delta G^{cd}}\right] ^{-1}=G^{ac}G^{db}+G^{da}G^{bc}
\label{freefluc}
\end{equation}
an eminently sensible result. Observe that the stochastic source does not vanish in this case, rather

\begin{equation}
\left\langle \kappa _{ab}\kappa _{cd}\right\rangle
=G_{ac}^{-1}G_{db}^{-1}+G_{da}^{-1}G_{bc}^{-1}  \label{freecor}
\end{equation}
However

\begin{equation}
\left( G^{-1}\right) _{ac}\sim -ic_{ac}\left( -\Box +m^2\right)
\label{freeinvprop}
\end{equation}
does vanish on mass - shell. Therefore, when we take the kinetic theory
limit, we shall find that for a free theory, there are no on - shell
fluctuations of the distribution function. 
For an interacting theory this is no longer the case.

The physical reason for this different behavior is that the evolution of the distribution function for an interacting theory is dissipative, and
therefore basic statistical mechanics considerations call for the presence
of fluctuations. Indeed it is this kind of consideration which led us to think about a Boltzmann - Langevin equation in the first place.

Conceptually, as mentioned earlier, the correct approach is to view the two point functions as an effectively open system separated from, yet interacting with, the hierarchy of higher correlation functions obeying the set of Schwinger-Dyson equations. The averaged effect of its interaction with an environment of slaved higher irreducible correlations brings about dissipation and the attending fluctuations give rise to the correlation noise. This is the conceptual basis
of our program laid out in \cite{CH95}, expounded in \cite{CH99} and completed in \cite{Cal09}. 


\subsection{Master Effective Action}

We motivated the existence of a fluctuation term in the Boltzmann equation for an interacting quantum field with the 2PI effective action which include the contribution of  both the one-particle distribution function and the two-point correlation function as independent variables, on the same footing. The generalization of this 2PIEA from n=2 to $n \rightarrow \infty$ is called the master effective action \cite{CH95}, a functional of the whole string of Green functions of the interacting field theory whose variation generates the Dyson - Schwinger hierarchy. To complete our journey we give here a formal construction of the master effective action. Further details of nPI techniques can be found in \cite{Berg,Carr}. 
 
We consider then a scalar field theory whose action
\begin{equation}
S[\Phi ]={\frac{1}{2}}S_2\Phi^2 +S_{int}[\Phi ]
\end{equation}
decomposes into a free part and an interaction part

\begin{equation}
S_{int}[\Phi ]=\sum_{n=3}^{\infty}{\frac{1}{n!}}S_n\Phi^n
\end{equation}
All fields are to be defined on a closed-time-path and DeWitt's condensed notation is adopted. We also use the shorthand

\begin{equation}
K_n\Phi^n\equiv\int~d^dx_1...d^dx_n~K_{na^1...a^n}(x_1,...x_n)
\Phi^{a^1}(x_1)...\Phi^{a^n}(x_n),
\end{equation}
where the kernel $K$ is assumed to be totally symmetric.

The `source action' is given by

\begin{equation}
J[\Phi ]=J_1\Phi +{\frac{1}{2}}J_2\Phi^2+J_{int}[\Phi ],
\end{equation}
where $J_{int}[\Phi ]$ contains the higher order sources

\begin{equation}
J_{int}[\Phi ]=\sum_{n=3}^{\infty}{\frac{1}{n!}}J_n\Phi^n.
\end{equation}
We define the generating functional

\begin{equation}
Z[\{J_n\}]=e^{iW[\{J_n\}]}=\int~D\Phi~e^{iS_t[\Phi ,\{J_n\}]},
\end{equation}
where

\begin{equation}
S_t[\Phi ,\{J_n\}]=J_1\Phi +{\frac{1}{2}}(S_2+J_2)\Phi^2+S_{int}[\Phi ]
+J_{int}[\Phi ].
\end{equation}
We shall also call

\begin{equation}
S_{int}[\Phi ]+J_{int}[\Phi ]=S_I.
\end{equation}

As it is well known, the Taylor expansion of $Z$ with respect to $J_1$
generates the expectation values of path - ordered products of fields

\begin{equation}
{\frac{\delta^n Z}{\delta J_{1a^1}(x_1)...\delta J_{1a^n}(x_n)}}= \langle
P\{\Phi^{a^1}(x_1)...\Phi^{a^n}(x_n)\}\rangle\equiv F_n^{a^1...a^n}
(x_1,...x_n)
\end{equation}
while the Taylor expansion of $W$ generates the `connected' Green functions (e.g., the `linked cluster theorem' of Haag)

\begin{equation}
{\frac{\delta^n W}{\delta J_{1a^1}(x_1)...\delta J_{1a^n}(x_n)}}= \langle
P\{\Phi^{a^1}(x_1)...\Phi^{a^n}(x_n)\}\rangle_{{\rm connected}}\equiv
C_n^{a^1...a^n} (x_1,...x_n).
\end{equation}
Comparing these last two equations, we find the rule connecting the $F$'s
with the $C$'s. First, we must decompose the ordered index set $(i_1,...i_n)$
($i_k=(x_k,a^k)$) into all possible clusters $P_n$. A cluster is a partition
of $(i_1,...i_n)$ into $N_{P_n}$ ordered subsets $p=(j_1,...j_r)$. Then

\begin{equation}
F_n^{i_1...i_n}=\sum_{P_n}\prod_{p}C_r^{j_1...j_r}.
\end{equation}
Now from the obvious identity

\begin{equation}
{\frac{\delta Z}{\delta J_{ni_1...i_n}}}\equiv \frac{1}{n!} {\frac{\delta^n Z%
}{\delta J_{i_1}...\delta J_{i_n}}}
\end{equation}
we obtain the chain of equations

\begin{equation}
{\frac{\delta W}{\delta J_{ni_1...i_n}}}\equiv {\frac{1}{n!}}
\sum_{P_n}\prod_{p}C_r^{j_1...j_r}.
\end{equation}

We can invert these equations to express the sources as functionals of the
connected Green functions, and define the master effective action (MEA) as
the full Legendre transform of the connected generating functional

\begin{equation}
\Gamma_{\infty}[\{C_r\}]=W[\{J_n\}]-\sum_n{\frac{1}{n!}}J_n\sum_{P_n}%
\prod_pC_r.
\end{equation}
The physical case corresponds to the absence of external sources, whereby

\begin{equation}
{\frac{\delta\Gamma_{\infty}[\{C_r\}]}{\delta C_s}}=0.
\end{equation}
This hierarchy of equations is equivalent to the Dyson- Schwinger series.

In summary, in the hierarchical representation of a closed quantum system a closed-time-path master ($n = \infty PI$) effective action is constructed,  whose functional variation generates the the Schwinger-Dyson  equations for the hierarchy of correlation functions.  When the hierarchy is truncated, one obtains the ordinary closed system of correlation functions up to a certain order, which obey a set of time-reversal invariant equations of motion.  But when the effect of the higher order correlation functions in the form of a correlation noise is included in one's consideration,   the dynamics of the lower order correlations shows dissipative features. These two procedures: \textit{`truncation'} versus \textit{'slaving'} (or causal factorization -- Boltzmann's molecular chaos assumption being a famous example) produce very different physical results. Truncation admits unitary evolution in the dynamical variables of both the lower sector and the higher sector of correlation functions, each can be regarded as a separate closed system; whereas slaving imparts dissipative dynamics in the lower-order correlation functions, as is well-known in the field-theory derivation of the Boltzmann equation by means of the 2PI effective action \cite{CH88}. Furthermore, the existence of a  fluctuation-dissipation relation for such an effectively open system mandates the existence of a stochastic term in the dynamical equation, representing the quantum fluctuations -- the correlation noise -- contributed by the higher order correlation functions. This is the origin of the stochastic Schwinger-Dyson equations,  in parallel to the classical Boltzmann-Langevin equation, which depicts both the dissipative and stochastic dynamics of correlation functions in interacting quantum field theory. 


\subsection{Decoherence by Correlation Noise and Correlation Entropy}

Having seen how correlation noise can be defined in a closed quantum system and how it imparts a stochastic component in the dissipative dynamics of the correlation functions, it is easy to see how decoherence comes about in a closed quantum system in a hierarchical representation. We mentioned earlier as the special role of the hydrodynamic variables which, thanks to the conservation laws they obey, habitually decohere and enable the emergence of relatively stable quasi-classical domains. Hierarchical representation offers the next best thing -- the correlation noise captures the effect of the higher correlations which acts like an environment to the lower correlations which induces decoherence. One can calculate the decoherence time scales by examining the multi-time scales of this environment which is inherited from the higher correlations,  or by directly solving the Boltzmann-Langevin equation.

A related physical quantity of fundamental importance is entropy. In the Boltzmann-BBGKY / Dyson-Schwinger framework Calzetta and Hu \cite{CH03} proposed the definition of a \textit{correlation entropy} of the nth order for an interacting quantum field. This is useful for addressing the thermalization of closed quantum systems and related quantum thermodynamic / kinetic theory issues. Using a CTP-2PI  effective action under a next-to-leading-order large N approximation they proved an H theorem for the correlation entropy of a quantum-mechanical O(N) model. 

Philosophically, the choice of hydrodynamical variables in the description of a closed quantum system of complex interactions is the key to the power and simplicity of a \textit{hydrodynamics theory}, while a hierarchical representation of a closed quantum system focusing on the nth order correlations, whose dynamics is captured by the nPI effective action, evokes the perspective of \textit{kinetic theory}.  Moving from a hydrodynamic depiction to a hierarchical representation enables one to zoom in from the macroscopic to the mescoscopic domains of a closed system.  \\

\noindent {\bf Acknowledgment} The two key ideas expounded here: the correlation noise, is contained  in prior work with Esteban Calzetta begun more than three decades ago, which he satisfactorily completed, while that of hierarchical coarse-graining, with Ryan Behunin, a decade ago and ongoing.  I'm thankful for their invaluable contributions which deepened my understanding of these fundamental issues of physics.

\end{document}